\documentclass[a4paper,openany, 12pt]{book}
\usepackage[text={15.5cm,23cm},centering]{geometry}

\newcommand\savemathcal[1]{%
  \expandafter\newsavebox\csname mc#1content\endcsname%
  \expandafter\savebox\csname mc#1content\endcsname{$\mathcal{#1}$}%
  \expandafter\newcommand\csname mc#1\endcsname{%
    \expandafter\usebox\expandafter{\csname mc#1content\endcsname}}%
}
\newcommand\altmathcal[1]{\csname mc#1\endcsname}
\savemathcal{Z}
\savemathcal{B}

\usepackage{mathptmx}
\usepackage{graphicx}
\usepackage{chapterbib} %remember to bibtex each chapter before final compilation
\usepackage{subfigure}
\usepackage{color}
\usepackage{amsmath}
\usepackage{amssymb}
\usepackage[colorlinks=true, linkcolor=blue]{hyperref}
\usepackage[english]{babel}  %For Vibor's institution characters

\title{The cosmic 21-cm revolution: charting the first billion years of our Universe}
\author{Andrei Mesinger}

\begin{document}
%\frontmatter
%\maketitle
%\tableofcontents

%\include{Mesinger/Preface}
%\include{Mesinger/Author}
%\include{Contributors} %only needed for edited books

%\mainmatter

%\include{chapter}
%\include{Furlanetto/chapter}
%\include{Mirocha/chapter}
%%\include{Pritchard/chapter}
%\include{Greig/chapter}
%\include{Bernardi/chapter}
%\include{Chapman_Jelic/chapter}
%\include{Greenhill_Subrahmanyan/chapter}
%\include{Trott_Pober/chapter}
\setcounter{chapter}{8}
\chapter{Future prospects}
\label{chapter:koopmans_bernardi}

\begin{bf}
L\'{e}on V. E. Koopmans (Kapteyn Astronomical Institute, University of Groningen),\\ 
Gianni Bernardi (INAF-IRA \& Rhodes University)\\
  
Abstract\\

\noindent This chapter addresses limitations to current 21-cm signal detection instruments, be it related to the instrument, environment, signal-processing or science, and what lies beyond the current horizon for 21-cm science, especially in the 2030s and beyond. We address how to overcome current challenges and drive the field forward, not only approaching a detection of the 21-cm signal but to a full characterisation of its parameter space, in particular, probing an increasingly larger volume of $k$-modes (spatially and in redshifts). We also will shortly touch upon the kinds of questions that could drive such future endeavours. 
\end{bf}

\section{What drives future 21-cm signal experiment?}

The past two decades have witnessed exciting advancements in the field of 21-cm Cosmology. Both theoretically and observationally great progress has been made, although a convincing detection of the 21-cm signal still has to be achieved both for the globally-averaged 21-cm signal as well as its spatial fluctuations. 
%We have observed the construction and operation of a vast number of ground-based interferometers (e.g. LOFAR, 21CMA, MWA, NenuFAR, LWA-OVRO, uGMRT, PAPER, [standard refs to these instruments]) and single-element receivers (e.g. EDGES, SARAS, BIGHORNS, PRIZM, SKYHI, [standard refs]), covering a wide range in their collecting area, core filling factor, field of view, frequency coverage, observational strategies (e.g. drift scan versus tracking), receiver technology (phased-array versus dishes), etc. 
We have observed the construction and operation of a vast number of ground-based interferometers  and single-element receivers, covering a wide range in terms of collecting area, core filling factor, field of view, frequency coverage, observational strategies (e.g. drift scan versus tracking) and receiver technology (phased-array versus dishes). 
Besides these instrument and technology developments driving the field forward, an enormous effort has been undertaken to develop much more refined flagging, calibration, imaging, foreground-removal, and 21-cm signal extraction methodologies (e.g., \cite{liu19}). These two tracks (instrument and signal processing) have gone hand in hand and have led to a steady progression and ever more stringent 21-signal limits (e.g., \cite{hera19}). The first confirmed 21-cm signal detection could be well in reach in the coming years. 

One of the most exciting and hotly-debated recent developments has been the announcement of the detection of the global 21-cm signal by the EDGES collaboration (\cite{bowman18}). Although confirmation of this claim is still needed, it shows that astrophysical effects (e.g. bright polarised foregrounds, ionospheric refraction, RFI mitigation) and instrumental challenges (e.g. chromatic leakage, band-pass structure, multi-path propagation, etc.) are controllable over nearly six orders of magnitude, and further improvements are still coming. 
These improvements in both instrument design, layout and interferometer versus single receiver technology, also inform each other and hybrid systems that are currently under construction (e.g. LEDA, NenuFAR).
Besides these ongoing experiments and observational programs, the next generation of instruments, of which HERA and the SKA (Section~\ref{sub_sec:hera} and \ref{sub_sec:ska}) are the most significant proponents, is now underway. Whereas currently instruments are limited to a ``statistical detection" of the 21-cm signal power spectra (opposed to the direct detection of the global signal; \cite{mesinger16}), mostly limited to the EoR due to the increasingly bright foregrounds and stronger ionospheric errors when moving to lower frequencies, these next-generation instruments aim not only to measure the 21-cm signal statistically but image it directly to the mK-level during the EoR and expand the redshift range to the Cosmic Dawn. This level of sensitivity requires a substantial increase in collecting area and filling factor (by a factor of about ten) over current instruments, thus stepping away for the experimental stage in which many active instruments find themselves. These new or upgraded systems are incorporating many of the lessons learned from past and ongoing efforts.   

In this chapter, we will touch upon some of these ongoing and forthcoming developments, although we will not describe each system in extreme detail. Several are already ongoing in terms of extensions of current instruments. 
%There is also the development of the SKA and HERA coming online fully in the early to later 2020s. 
Finally we shortly contemplate what lies beyond the current 2030 horizon, in particular instrumentation that can expand the currently envisioned science scope of the next-generation instruments and might require significantly larger collecting areas ($\gg$1~km$^2$) and deployed in space (including the lunar environment or surface) to allow one to escape the limits set by the ionosphere and human-made interference at low frequencies. This would allow them to observe the holy grail of high-redshift 21-cm Cosmology, being the era called the {\sl Dark Ages}, which allows a direct probe of questions posed by fundamental physics, inaccessible via any other way than except the 21-cm signal (\cite{koopmans19}).

\subsection{Limits of current 21-cm signal observations}

As shown in \cite{mellema13} and \cite{koopmans15}, core collecting area, its filling factor, and the field of view (FoV) of the instrument primarily drive the statistical sensitivity of an interferometer to the 21-cm signal power spectrum and its imaging capabilities. For direct imaging of the 21-cm signal or fluctuations on small scales, the field of view (FoV) is less critical since cosmic variance might not be the driving factor. A limited FoV will increase the sample variance on the large scales, however, in power-spectrum measurements. Besides the sensitivity to a given 21-cm signal mode in the presence of thermal noise, there are many other instrumental limitations. Some of these limitations can be kept under control to some extent but some cannot, and thus have to be avoided (e.g., by choosing the proper location or controllable experimental set-up) or mitigated (i.e., correct for errors in the data in real-time or during post-processing).
Below we summarize some of the issues that are currently considered as limiting 21-cm experiments to reach the thermal noise:
\begin{itemize}
%\noindent 
\item (a) Collecting area: Although collecting area is one of the driving factors in sensitivity, it also drives hardware costs, especially if the receiver systems being correlated are small and many are needed to reach the required collecting area. Lack of collecting area can also be compensated by increasing the field of view of a system such that more measurements are made of the same $uv$-cells, and more $uv$-cells are sampled, increasing power spectrum sensitivity. This drives up correlator costs, however, so a careful balance needs to be struck between the difference requirements; 
\item (b) Filling factor: Placing receivers in a smaller core-area, even for fixed collecting area and field of view, increases the sensitivity of the instrument for the simple reason that more independent visibilities are measured per $uv$-cell (i.e. $k$-mode). Since the 21-cm signal of interest is the same per visibility, whereas thermal noise is not, a higher filling factor rapidly increases the power-spectrum sensitivity; 
\item (c) Field of View: A larger field of view, in general, is related to a smaller receiver element, and more elements to be correlated. For the aforementioned reasons, this increases the number of independent $uv$-cells and independent 21-cm signal $k$-modes, thereby decreasing power spectrum errors; 
\item (d) Frequency Coverage: Whereas maximizing the frequency coverage from the EoR to the Cosmic Dawn, and even into the Dark Ages (i.e., $z\sim 6 - 200$) would be optimal, in practice this frequency range is split up in smaller bands. These bands are generally limited in their spectral resolution, some channels are lost to RFI (e.g. FM band, DAB/DVB; see below) and in some cases not even covered (e.g. below the ionospheric cutoff which limits observations of the Dark Ages). These instrumental and environmental effects have led to the development of new wide-band receivers, the deployment of instruments in remote locations to avoid RFI, or even in space where they are not affected by the ionosphere (see below); 
\item (e) $uv$-coverage: This has already been discussed above, and in general is driven by the number of correlated receivers and the density of the core. The limiting factor is often costs of the electronics, the correlators and data storage. However, in some cases longer baselines are necessary to calibrate the instruments in case it is a highly non-redundant array; 

\item (f) (Polarised) Foregrounds: Foreground emission is a significant complicating factor in a 21-cm signal experiment, not only because they are bright but also because they are partly polarised and have spatial structure. These two effects couple to the ionosphere and the chromatic and polarised nature of the instrument itself even in the absence of any errors, and cause leakage terms from the foregrounds into the 21-cm signal; 
\item (g) Instrumental effects: Instrument are not perfect. Receivers need amplifiers to increase the weak electromagnetic signals into a measurable voltage that can be digitized. These low-noise amplifiers are not 100\% stable and can cause both amplitude and phase errors in the visibility data after correlation. When various receivers and amplifiers take part in station-beamforming, these errors become direction-dependent. Secondly, cross-dipole receivers or receivers that measure circular polarisation partly mix Stokes $I$, $Q$, $U$ and $V$ power. The reason is that antennas see radiation coming from different projections of the sky, leading to instrumental polarisation. So even if the instrument is nearly perfect, these effects are impossible to avoid and if the sky is partly-polarised, this power contaminates the 21-cm power spectrum; 
\item (h) Signal processing: One final difficulty is inherent to the data processing. Since processing affects the data via RFI excision, gain calibration, foreground-removal, etc.; it can both remove and enhance signals of interest from the data, including the 21-cm signal itself. Processing, if not done very precisely and accurately, can thus lead to a severe 21-cm signal bias (in general suppression).  
\end{itemize} 

\noindent Besides these limiting factors, mostly instrument-related, two crucial factors are harder to control and will ultimately drive the designs of large-scale future low frequency instruments towards space.\\

\noindent {\bf (1) Radio Frequency Interference:} Radio frequency interference (RFI) is becoming an increasingly more critical problem for low-frequency telescopes, as human-made signals from, e.g., transmitters, vehicles, mobile phones, satellites, aeroplanes, are now occupying many of the frequency bands previously clean (\cite{offringa13}, \cite{offringa15}). This increasing RFI occupancy motivates the next generation of instruments to be built in remote desert environments such as the Karroo in South Africa and Western Australia. Going to space is another, albeit expensive option. Just above the Earth, however, any receiver would see a much larger number of transmitters, making the problem worse. Even at a distance of the moon, an RFI suppression of eight orders of magnitude in power would be needed to mitigate it to a level that the 21-cm signal from the CD can be observed. In a lunar orbit, the moon would shield the receivers from Earth and also from solar radiation for a fraction of the time, creating an "RFI-free" cone.
%. Recent activity on the lunar far-side (see Section~\ref{sec:space_base_instruments}), however, might also jeopardize this as well. 
A receiver on the fare-side of the moon itself might be shielded even more. 
%(if far removed from surface activities), although charged particles on the lunar surface might impact such an instrument as well. Despite this, 
Future low-frequency instruments will likely be in space to at least partly mitigate the worsening RFI situation on Earth. \\

\noindent {\bf (2) Ionosphere:} The ionosphere causes both phase and amplitude fluctuations in the received electro-magnetic signal, which increase in strength toward lower frequencies, and maximal near the plasma frequency cutoff of the ionosphere (around $5-10$~MHz). The ionosphere has restricted most instruments to frequencies above $30-50$~MHz, up to the Cosmic Dawn. Observations of the Dark Ages will be extremely hard, especially in the presence of very bright foreground that couple to the ionosphere (\cite{vedantham15}, \cite{vedantham16}). Reaching the Dark Age 21-cm signal, therefore, requires space-based instruments. \\
%Even the early Cosmic Dawn around $z \sim 30$ is effectively out of reach because of the ionosphere even forthcoming instrument (see Section~\ref{sec_ground_base_interferometers}) may be cover those frequency bands.\\

\noindent Whereas these technical and environmental reasons will ultimately drive instruments to become ever larger and be built on remote places, or even in space, what do we hope such future instruments will tell us?

\subsection{What will drive future 21-cm experiments}

Future experiments will primarily be driven by increasing sensitivity in 21-cm signal regimes already explored, but also by exploring new regimes in redshift and spatial scales. The former has fundamentally driven the development of the high-sensitivity SKA and HERA telescopes, with their distinct approach in sensitivity increase in different parameter space regions (see Section~\ref{sec_ground_base_interferometers}). Below we will shortly discuss where future instruments will represent a leap forward with respect to current instruments (assuming the latter will detect the 21-cm signal in the coming years). 

\begin{itemize}
  
\item {\bf Smaller spatial scales:} Most current instruments are limited to rather a small range (less than one decade) power spectrum $k$-modes where sufficient signal-to-noise can be reached for a detection. The reason is that shorter baselines add coherently for a more extended integration time per $uv$-cell and these are most sensitive to the larger spatial cases (except in the frequency direction). To observe larger $k$-modes (or smaller three dimensional spatial scales), instruments in general need much more collecting area, one of the SKA drivers. HERA aims to reach higher sensitivity by  increasing collecting area, its field of view and its the filling factor, but at the cost of having redundant $uv$-sampling. The latter makes direct imaging much harder, especially on smaller spatial scales, also posing limitation to calibration strategies. 
%However, small spatial scales can also be probed via the power spectrum since they are sampled both in the spatial domain as well as in the frequency domain (since frequency corresponds to redshift and hence to comoving distance). SKA aim for minimum redundancy and henced needs more collecting area than HERA, but its imaging capabilities are far superior.
    
\item {\bf Direct imaging:} Direct imaging of the 21-cm signal becomes possible with larger collecting area and sensitivity on all spatial scales. One of the highest-priority science drivers for the SKA is, therefore, the direct imaging  of the 21-cm signal throughout the EoR on scales larger than ten arcminute (see Section~\ref{sub_sec:ska}). Ionized bubbles can be imaged on even smaller scales since their contrast to the globally-averaged 21-cm signal is very large (about 30~mK versus fluctuations of only a few mK around the bubbles). 

\item {\bf Higher redshifts:} A third driver behind the sensitivity increase is that it enables observations at even higher redshifts, e.g., in the Comic Dawn or even the Dark Ages. At higher redshift, the foregrounds are much brighter, increasing the overall system temperature and hence the thermal noise. Integration times therefore rapidly increases with redshift, and observations of the Cosmic Dawn are the territory of the upcoming interferometric arrays.
\end{itemize}

\noindent In short, whereas present-day instruments aim for the first detection at several (lower) redshifts and over a limited range of spatial scales, the next generation of instruments will aim to expand these parameter-spaces, but also do direct imaging rather than summarising the signal in some statistics (e.g. the power-spectrum). In Section X, these next-generation instruments will be discussed, where we not only limit ourselves to SKA and HERA but also shortly touch upon extensions to current instruments and their main science drivers.

\section{Ground-based interferometers}
\label{sec_ground_base_interferometers}

In this section, we review the status of those 21~cm ground-based interferometers that are under construction, have been upgraded or will be constructed shortly.

\subsection{The Square Kilometre Array -- SKA1\&2}
\label{sub_sec:ska}

The Square Kilometre Array (SKA hereafter) is a global effort by a consortium of member-state countries\footnote{www.skatelescope.org} that will consist of at least two entirely different arrays: a mid-frequency array  (SKA-mid) to be built in the Karoo desert of South Africa, and a low frequency array (SKA-low) to be built in Western Australia, both located in radio-quiet zones where human-made RFI is very limited. 
%The site of SKA-mid currently hosts MEERKAT, which will become part of the SKA, and HERA, whereas the site in Western Australia hosts both the MWA and ASKAP. 
Since only SKA-low will cover the redshifts of the Cosmic Dawn and Epoch of reionization ($50-200$~MHz, or $6 < z < 27.4$), here, we focus only on this instrument. 
\begin{figure}[]
\begin{center}
\includegraphics[width=0.6\textwidth]{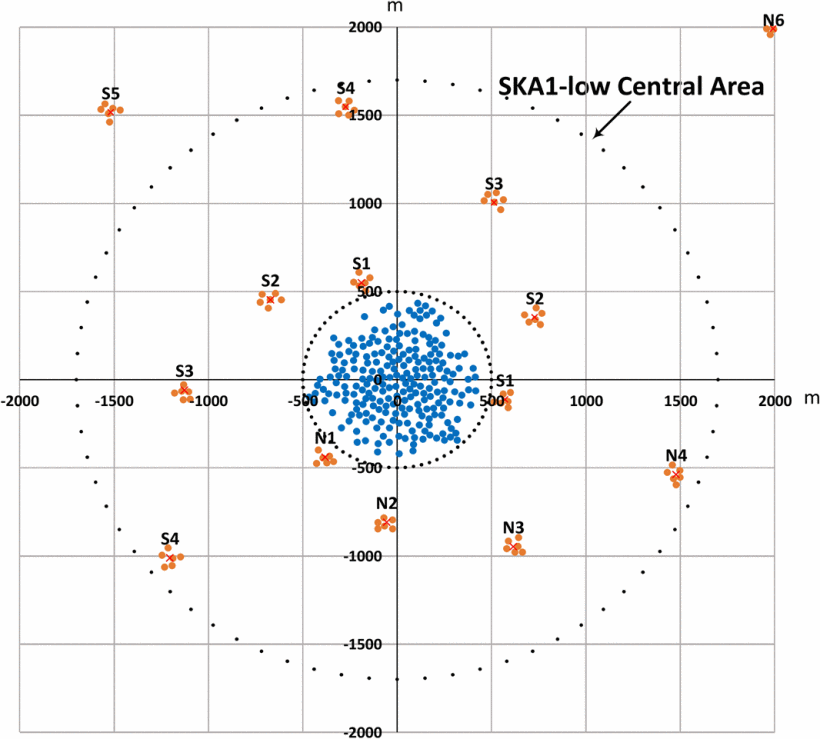}
\end{center}
\caption{SKA1-low core layout, showing the station positions in the inner few km, which are most sensitive to the 21-cm signal (\cite{dewdney09}, \cite{waterson16}).}
\label{fig:fig_ska_layout}
\end{figure}
SKA-low has a design similar to LOFAR (\cite{mellema13}, \cite{koopmans15}), but has some distinct differences as well, primarily related to the receiver design. SKA-low aims to have 512 stations in Phase 1 (denoted here by SKA1-low), having 256 log-periodic cross-dipole receivers that cover the full frequency band and are semi-randomly spread inside a 40~m diameter circle. The requirement of a high-gain receiver over the full spectral band limits the field of view to a cone with an opening angle of about 90 degrees centred on the zenith, but one that maximizes forward gain. The current receiver design also aims for a spectrally smooth bandpass, something rather difficult to realize for a wide-field and wide-band receiver (refs).
About 212 stations will be placed inside a central core of about 600~m (Figure~\ref{fig:fig_ska_layout}), making it about eight times more sensitive than LOFAR. The remaining stations will be distributed along three arms that "spiral" outward up to about 65~km, in the current design.  The long baselines will enable proper direction-dependent gain calibration of the system. The SKA observational and calibration strategy leverages on minimal redundancy to reduce the point spread function side-lobes, improve imaging fidelity and sky-based calibration capabilities. 
%It remains to be seen which strategy is the most optimal, although one should keep in mind that SKA-low is also built for other science cases, whereas HERA is a pure 21-cm signal experiment and does not have to cater to other science interests. 
%
\begin{figure}[]
\begin{center}
\includegraphics[width=\textwidth]{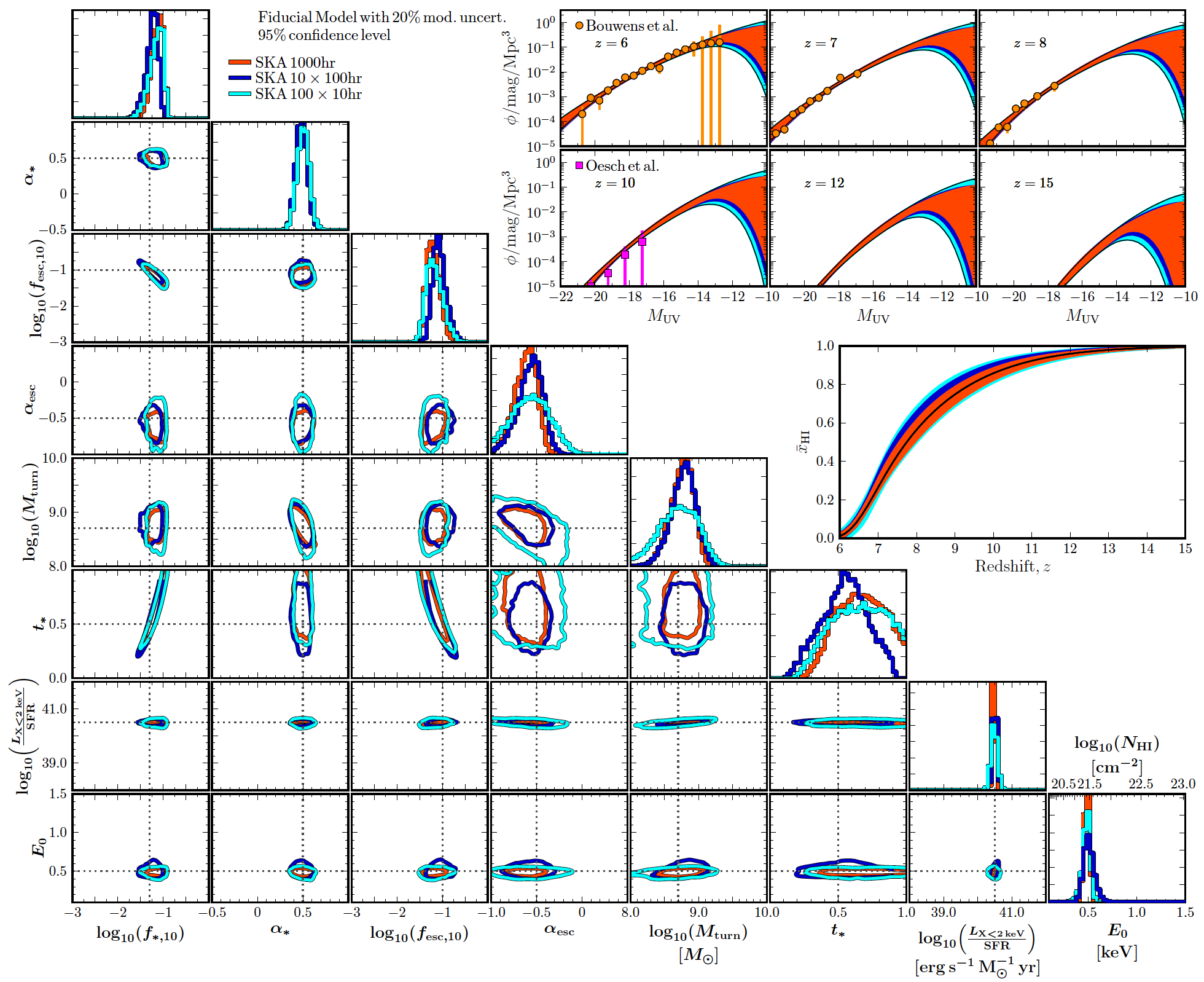}
\end{center}
\caption{Astrophysical parameter constraints for a standard 21-cm signal model using 1000~h of observations with SKA1-low (\cite{greig19}).}
\label{fig:fig_ska_astroph}
\end{figure}

The sheer collecting area (0.4~km$^2$) and instantaneous (300~MHz, or 150~MHz when split in dual-beam mode) bandwidth make SKA1-low the premier instrument in the late 2020s to directly image the 21-cm signal during the EoR from $6 < z < 12$, covering the 21-cm signal power-spectrum in the range of roughly $ 0.02 < k < 1 $~cMpc$^{-1}$ (depending on redshift), and push power-spectrum measurements over more limited $k$-ranges deep in to the Cosmic Dawn, up to $z \sim 20$ (or even more). 
%In particular the former will make this a transformational instrument. For example, observing 
Observations of individual ionized structures will enable cross-correlations with many other instruments that aim to look for the sources of reionization (e.g., JWST, ALMA, SPICA). Figure~\ref{fig:fig_ska_astroph} shows an example of how well SKA-low will constrain reionization parameters (\cite{greig19}). Currently, SKA is planned to be operational around 2028, although early science is foreseen several years before that.
Finally, ideas for a far-future ($\gg 2030$) upgrade to SKA2-low have already been developed (e.g., \cite{koopmans15}), which nominally foresees an increase in collecting area by a factor of about four. This could allow more detailed imaging due to lower thermal noise, and increase angular resolution also at higher redshifts.
%but also lower point spread function side-lobes, and hence improved image fidelity. Longer baselines also help calibration, foreground subtraction and enable even higher redshifts and broader ranges of $k-$ modes to be investigated in terms of 21-cm power spectrum measurements.

\subsection{The Hydrogen Epoch of Reionization Array -- HERA}
\label{sub_sec:hera}
\begin{figure}[]
\begin{center}
\includegraphics[width=1.\textwidth]{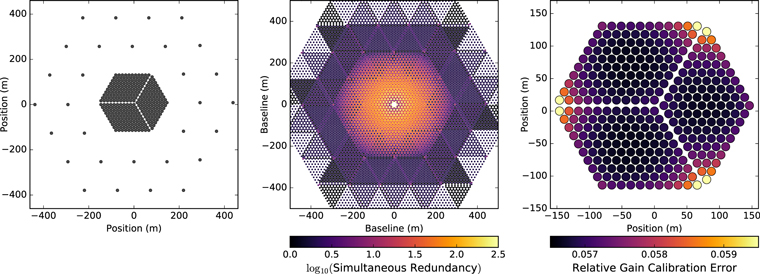}
\end{center}
\caption{The HERA layout (left panel): 320~dishes are located in the hexagonal core and 30 more outrigger dishes are planned to be deployed out to a maximum baselines of $\sim 800$~m to improve angular resolution and imaging capabilities. The core is split in three sectors that are displaced from each other by a fraction of the dish diameter (see \cite{dillon16} for a detailed discussion. The split core provides a significantly improved instantaneous $uv$ coverage (central panel) whilst retaining high redundancy. The right panel shows the expected relative antenna gain errors after using redundant calibration (from \cite{dillon16}).}
\label{fig:fig_hera}
\end{figure}
\begin{figure}[]
\begin{center}
\includegraphics[width=0.6\textwidth]{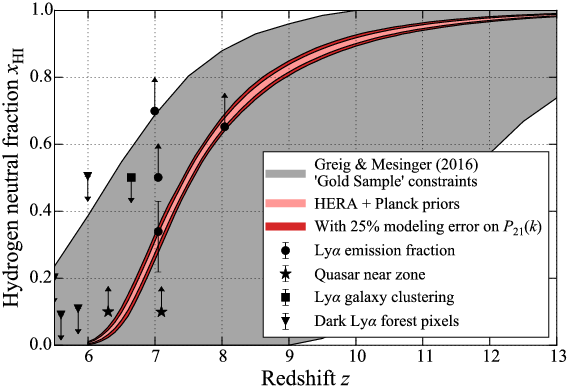}
\end{center}
\caption{95\% confidence region on the Hydrogen neutral fraction $X_{\rm HI}$ (grey, from \cite{greig17}). The inclusion of HERA measurements leads to a dramatic improvement in the constraints (red and pink areas, \cite{liu16b}). Constraints from other reionization probes are shown as well (see \cite{deboer17} for a detailed description).}
\label{fig:fig_hera_ion_hist}
\end{figure}

The Hydrogen Epoch of Reionization Array (HERA, \cite{deboer17}) is an array currently under construction in the Karoo reserve area in South Africa - following the decommissioning of the PAPER experiment (see Chapters~5 and 8 in this book for an overview of PAPER). HERA is built following the approach used for PAPER: a highly redundant array to maximize the sensitivity on a number of power spectrum modes measured using the avoidance approach. In order to increase the sensitivity with respect to PAPER, it employs 14~m diameter non steerable dishes that, in the final configuration, will be densely packed in a highly redundant hexagonal array configuration of $\sim 350$~m diameter (see Figure~\ref{fig:fig_hera}). 
HERA main goal is to measure the 21~cm power spectrum in the $6 < z < 12 $ range with high significance in the $0.2 < k < 0.4$~Mpc$^{-1}$ range (\cite{pober14}, providing a full characterization of the evolution of the neutral Hydrogen fraction of the intergalactic medium (Figure~\ref{fig:fig_hera_ion_hist}).
Given the high redundant configuration, imaging tomography will remain challenging for HERA and likely the goal of a future generation experiment. As foreground modeling and characterization will also be limited because of redundancy and the coarse angular resolution, a significant effort was dedicated to keep the instrumental response from corrupting the intrinsically smooth foreground spectra and to accurately model it (\cite{neben16}, \cite{ewallwice16}, \cite{thyagarajan16}, \cite{patra18}). An alternative approach to redundant calibration is to apply foreground avoidance using closure phase quantities from antenna triads (\cite{thyagarajan18}): closure phase are insensitive to errors in direction independent interferometric calibration and, therefore, directly bypass the requirement of an accurate spectral calibration (see Chapter~5 in this book for an overview of calibration of 21~cm observations). A preliminary analysis of HERA closure phases seem to confirm these premises (\cite{carilli18}).
HERA is currently under construction, with more than 200 dishes deployed, and 21~cm observations are currently being analyzed. New feeds that extend the sensitivity to the 50-250~MHz range are currently deployed for testing in order to enable observations in the $12 < z < 35$ range (the Cosmic Dawn) and probe the nature of the first luminous sources and their impact on the thermal history of the intergalacic medium.

\subsection{The Large aperture Experiment to detect the Dark Ages -- LEDA}
\label{section:leda_pspec}
\begin{figure}[]
\begin{center}
\includegraphics[width=0.6\textwidth]{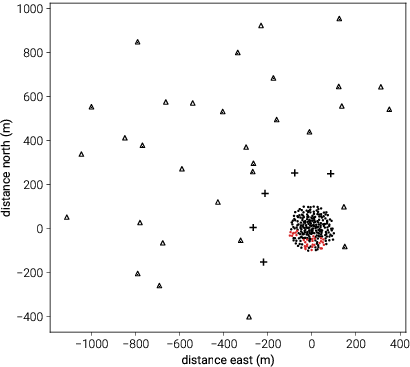}
\end{center}
\caption{LEDA antenna layout: the dense core is surrounded by 32 dipoles in order to provide an exceptionally good instantaneous $uv$ coverage (from \cite{eastwood18}).}
\label{fig:fig_leda}
\end{figure}
The Large aperture Experiment to detect the Dark Ages (LEDA, \cite{bernardi15}, \cite{kocz15}) is located at the Owens Valley Radio Observatory, California. It operates in the 30-88~MHz frequency range, corresponding to $15 < z < 46$, seeking to detect the 21~cm signal from the Cosmic Dawn.   
The array layout consists of 251 dipoles pseudo randomly deployed within a 200~m diameter core, 23 dipoles are added out to a maximum 1.5~km baseline (see Figure~\ref{fig:fig_leda}). Five additional outrigger dipoles are custom-equipped to measure the global 21~cm signal via individual custom-built dipoles (see Section~\ref{leda_global}).
The very dense core provides exceptional brightness sensitivity and a point spread function with very low sidelobes. The outrigger dipoles improve the angular resolution that helps to identify calibration sources and lower the confusion level. As the dipoles are individually correlated, visibilities have contributions from all-sky emission, particularly from Galactic diffuse emission - given the number of short baselines - and with significant ionospheric-induced refraction and scintillation. Despite these challenges, \cite{eastwood18} generated the first high quality all-sky foreground maps. 
The LEDA approach to measure the 21~cm signal can be versatile, allowing to image and subtract foregrounds (\cite{eastwood18}) but also to avoid them (similar to \cite{beardsley16}). \cite{eastwood19} analyzed 20~hours of LEDA data calibrated using a compact source sky model and filtering foregrounds by using their statistical properties in way similar to \cite{dillon14} and \cite{trott16}. They reported an initial $10^8$~mK$^2$ upper limit on the 21~cm power spectrum at $z = 18.4$. Several hundreds of hours of observations have been collected now and will be the focus of future analysis towards the detection of the power spectrum from the Cosmic Dawn and an independent confirmation of the reported detection by \cite{bowman18}.

\subsection{The Low Frequency Array 2.0 -- LOFAR2.0}
\label{section:lofar}

\begin{figure}[t]
\begin{center}
\includegraphics[width=0.8\textwidth]{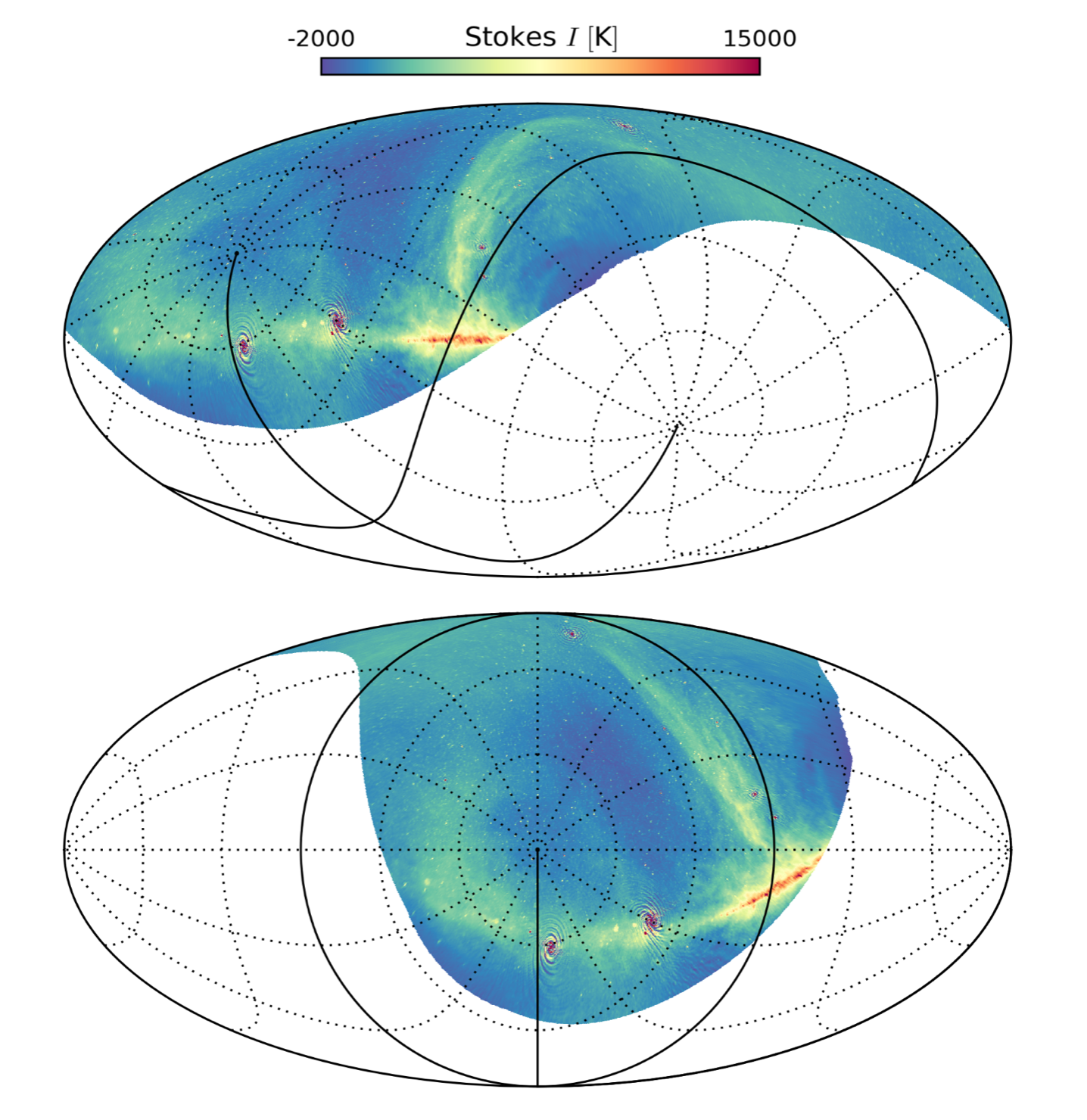}
\end{center}
%\caption{The first all-sky image of the Northern sky taken with the AARTFAAC-LBA system, using 576 dipoles on the inner 12 LOFAR-LBA stations. Figure credit: B. Gehlot.}
\caption{The first all-sky image of the Northern sky taken with the AARTFAAC-LBA system, using 576 dipoles on the inner 12 LOFAR-LBA stations.}
\label{fig:fig_AARTFAAC}
\end{figure}
LOFAR (\cite{vanhaarlem13}) is already one of the most sensitive arrays to detect the 21-cm signal during the EoR, although its sensitivity is still limited in the Cosmic Dawn redshift/frequency-range. This is due to the limited effective Low-Band Antenna (LBA) collecting area, and the system temperature at low frequencies being much larger. Furthermore, the LBA dipoles are rather narrow-band, and the gains sharply peak around 60~MHz, dropping rapidly at frequencies away from the resonance. 
%Similarly, at low frequencies,  ionospheric phase fluctuations increase rapidly making these data harder to calibrate and harder to reach the thermal noise. To mitigate both problems, 
LOFAR will undergo upgrade in the coming years in order to improve its sensitivity and capability to calibrate the ionospheric distortions. Sensitivity will be increased by connecting the 48 dipoles in each LBA station that are not connected to the acquisition system because of budget limitations.

%Firstly, half of the 96 dipoles currently in each LBA station is not connected for cost reasons (the cost of an LBA dipole is an order of magnitude lower than the electronics needed to connect it to the overall system), despite being in the field. During the upgrade of LOFAR to LOFAR2 [https://www.astron.nl/lofar-20-newsletters-home], these 48 LBA-dipoles per station will be connected to the beam-former, effectively doubling the LOFAR-LBA collecting area. This increases LOFAR's sensitivity but also its ability to calibrate on more (fainter) sources, thereby improving ionospheric corrections. Connected to this, currently, only observations can be done in either HBA or LBA mode. 

Moreover, LOFAR2 will enable simultaneous observations with the HBA and LBA systems, such that the more sensitive HBA system can be used to gain-calibrate the system, including (direction-dependent) ionospheric corrections. Finally, LOFAR2 will in a later stage also enable HBA observations with a dual analogue tile-beam formation, enabling multiple target fields anywhere on the visible sky (not just limited to multiple beams inside the HBA-tile beam, as is currently the case). Each of these upgrades improves the thermal-noise sensitivity to the 21-cm signal and the system calibratability. The first step in this process 
%GPU-based correlator, COBALT, to COBALT2, 
was recently taken with the installation of a GPU-based correlator.  

%\subsubsection{Amsterdam-ASTRON Radio Transients Facility And Analysis Center -- AARTFAAC}

\begin{itemize}
\item {\bf Amsterdam-ASTRON Radio Transients Facility And Analysis Center -- AARTFAAC}.\\
Whereas LOFAR mainly operates in beam-formed mode, where dipoles or tiles are phased-up in a given direction, the AARTFAAC system\footnote{www.aartfaac.org} currently enables all 576 LBA or HBA dipoles/tiles of the inner 12 stations to be cross-correlated, using two physical correlators, although currently only over a very limited 3.1-MHz bandwidth with 60-kHz resolution. This operational mode increases the field of view by a factor of about 25 for the HBA system and to all-sky for the LBA system (Figure~\ref{fig:fig_AARTFAAC}), improving the power-spectrum sensitivity by a factor of about five or more per unity bandwidth (since both the collecting area and filling factor remain similar and long baselines do not add sensitivity to the 21-cm signal). AARTFAAC is currently already being used to target, for example, the 21-cm signal in the Cosmic Dawn with the goal to interferometrically confirm the recently reported global signal detection (\cite{bowman18}). A system upgrade where all dipoles/tiles are fully correlated for 24 stations is envisioned over the full LBA and/or HBA bandwidth. 
\end{itemize}

\subsection{The Murchison Widefield Array phase II}

The Murchison Widefield Array (MWA) is located in Western Australia, operating in the $80-200$~MHz range. In its initial phase (phase I, \cite{tingay13}) it consists of 128 stations deployed in a pseudo random configuration out to a $\sim 3$~km baseline. Each station includes 16 bow-tie dipoles arranged in a regular $4 \times 4$ square grid, 5~m wide. Like LOFAR, MWA is a general purpose instrument, although one of its main science drivers is the measurement of the 21~cm signal from the EoR. It recently underwent an expansion (termed ``phase II", \cite{wayth18}) where the number of stations was doubled: 72 new stations were placed in two highly redundant hexagons next to the array centre and 56 stations outside the phase~I array to extend the maximum baseline up $\sim 5$~km. Only 128 stations can still be instantaneously correlated, resulting in two different configurations: a compact configuration that includes the two redundant hexagons and the phase~I compact core, and an extended configuration that excludes the two redundant hexagons and includes stations out to the longest baselines (see Chapter~8 in this book for an illustration of both configurations).

\begin{figure}[t]
\begin{center}
\includegraphics[width=0.8\textwidth]{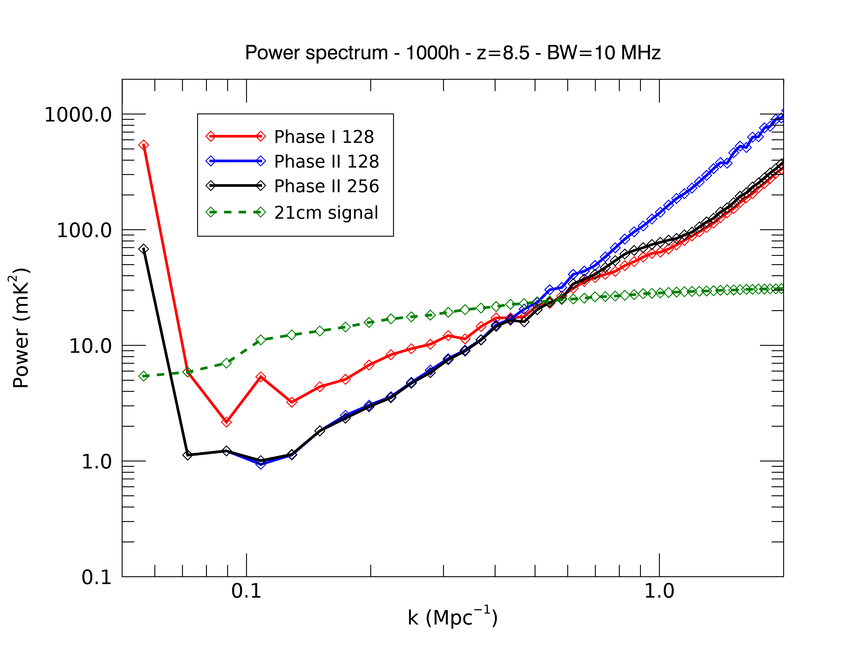}
\end{center}
\caption{Fiducial 21~cm power spectrum model at $z = 8.5$ with associated noise levels from Phase I and Phase II arrays with a 1000~hour observation. ``Phase II 256" shows the result from a future MWA upgrade where all 256 tiles are correlated simultaneously (from \cite{wayth18}).}
\label{fig:fig_mwa_phaseII_pspec}
\end{figure}

The MWA phase~II array is therefore a fairly flexible instrument: its compact, redundant configuration, is optimized for EoR power spectrum observations following a strategy similar to HERA, i.e. leveraging upon redundant calibration schemes (\cite{li18}) and improving the sensitivity by a factor of four with respect to phase~I, leading to a $\sim 10\sigma$ detection of the fiducial 21~cm power spectrum at $k \sim 0.1$~Mpc$^{-1}$ (Figure~\ref{fig:fig_mwa_phaseII_pspec}).

In its extended configuration it has an exceptionally good instantaneous $uv$ coverage (due to the high number of stations instantaneously correlated) with improved foreground modeling thanks to the increased angular resolution from phase~I.

\subsection{New Extension in Nancay Upgrading LOFAR -- NenuFAR}

\begin{figure}[t]
\begin{center}
\includegraphics[width=0.8\textwidth]{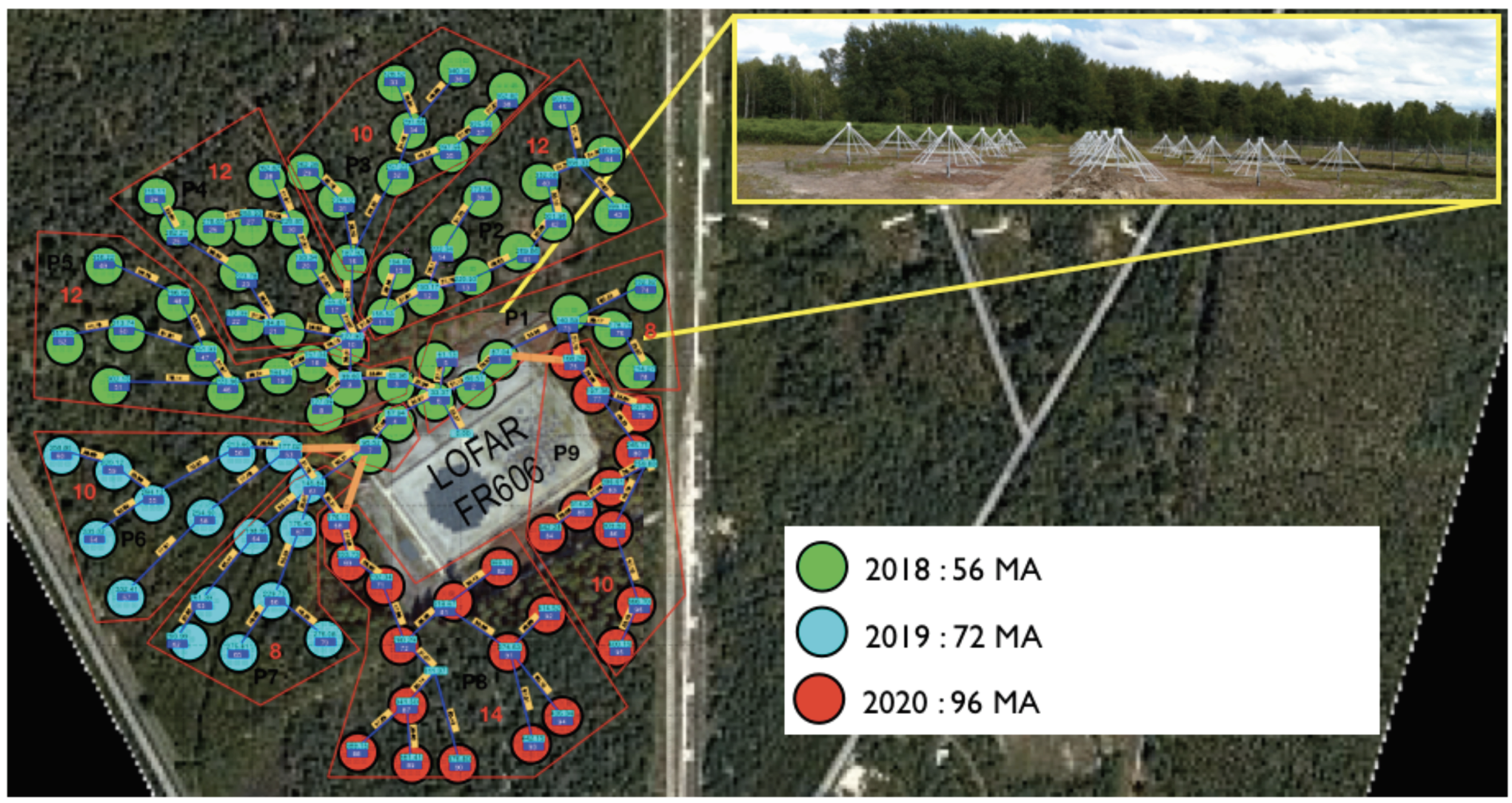}
\end{center}
\caption{The current and planned layout of NenuFAR in Nancay, France. By late 2019, around 80 mini-arrays (stations) will be in place and operating, to be expanded to the full 96 stations in 2020 (and beyond). A new GPU-based correlator will also be installed (\cite{french_ska_white_book}).}
\label{fig:fig_NenuFAR}
\end{figure}

Another novel array currently in its roll-out and early-science phase is NenuFAR (Figure~\ref{fig:fig_NenuFAR}; \cite{french_ska_white_book}). Whereas initially envisioned as an extremely sensitive beam-formed system operating in  the $(10)30-85$~MHz, the development of a relatively cheap GPU-based correlator for LOFAR will enable NenuFAR to correlate all envisioned 96 mini-arrays, each consisting of 19 LEDA-like dipoles, over the full frequency band. 
%if high spectral resolution (64 channels of $\sim$3\,kHz each per sub-band of 195\,kHz). 
The field of view of NenuFAR is about 20 degrees at 60~MHz, and, inside the 400~m core, the filling factor reaches order unity at 35~MHz, and $\sim 0.25$ at 60~MHz, which makes it extremely sensitive to low-surface brightness structures. Currently, 56 stations are in place inside a core of about 400~m diameter.
%, which can be correlated only over 3.1\,MHz of 16 sub-bands (each of 195\,kHz). 
By late 2019, however, 80 stations will be in place, six of which  will be placed further out over an $\sim 2.5$~km diameter area. The correlator is already being installed and will also be operational by late 2019. The final goal is to have 96 mini-arrays in place, enabling maximum use of the system. This will make NenuFAR one of the most sensitive 21-cm signal arrays in the world, in principle able to reach the standard predicted 21-cm signal in the Cosmic Dawn redshift range (\cite{mesinger16}). 
%A large observational (key-science) program has started in 2019 to enable this over the coming years.     

\section{Global Signal Experiments}

In this Section we briefly review the status of ongoing global signal experiments.  
%(see Chapter~7 for a more detailed discussion about global signal observations).

\subsection{The Experiment to Detect the Global EoR Signature -- EDGES}
%\begin{figure}[]
%\begin{center}
%\includegraphics[width=1.\textwidth]{Koopmans_Bernardi/edges_trough}
%\end{center}
%\caption{EDGES}
%\label{fig:fig_edges}
%\end{figure}
The Experiment to Detect the Global EoR Signature (EDGES, \cite{bowman08}) currently operates in two frequency bands: the $90-200$~MHz (high) band in order to constrain the evolution of the neutral fraction throughout reionization, and the $50-100$~MHz (low) band, in order to measure the expected heating of the intergalactic medium from primordial sources. 
The EDGES experiment has been pioneering techniques to accurately model all the various instrumental components in order to carefully control systematics effects (\cite{bowman08}, \cite{monsalve17a}). Observations in the high band have constrained the duration of reionization $\Delta z$ to be longer than $\Delta z >  1$ and started to constrain some properties of the first galaxies (\cite{monsalve17b}, \cite{monsalve18}). In the low band, \cite{bowman18} reported the surprising detection of an absorption trough twice deeper than the most extreme models, posing a serious challenge to its interpretation - assuming it is of cosmological origin. 
In the light of this anomalous signal, the EDGES team is deploying a new dipole antenna tuned in size to simultaneously observe the $60-160$~MHz range (i.e. $\sim 25\%$ smaller than the low band antenna) and confirm the results in the low band. A further upgrade of the EDGES experiment with a more portable antenna that includes the electronics is under consideration for deployment in a quiet radio frequency environment in Oregon, USA.

\subsection{The Large aperture Experiment to detect the Dark Ages -- LEDA (global signal)}
\label{leda_global}
%\begin{figure}[]
%\begin{center}
%\includegraphics[width=1.\textwidth]{Koopmans_Bernardi/leda_dipole}
%\end{center}
%\caption{LEDA dipole}
%\label{fig:fig_leda_dipole}
%\end{figure}
As mentioned in Section~\ref{section:leda_pspec}, LEDA includes a few custom-equipped dipoles to measure the global signal (\cite{price18}). Initial observations were used to validate the end-to-end acquisition system and data analysis, leading to a 890~mK upper limit on the global signal amplitude in the $13.2 < z < 27.4$ range at the 95\% confidence level (\cite{bernardi16}). A series of upgrades have been implemented since the early system: filters with a sharper roll-off were installed in order to improve RFI rejection and extend the observing band up to 87.5~MHz; the noise diode stability was improved and a system to measure the ambient temperature was installed on the dipoles. The receiver seems to show the necessary stability to measure the global signal, however, other sources of systematic effects related to the antenna gain pattern remain less well known and are the subject of ongoing modeling and investigation.
About 100~hours of observations were taken with the upgraded system and are currently being analyzed.
%
%{\bf (GB: this section may be removed if already included in Chapter~7, I cannot see that chapter yet...)}

\subsection{Shaped Antennas to measure the background RAdio Spectrum -- SARAS}

The Shaped Antennas to measure the background RAdio Spectrum (SARAS) represent a progression of radiometers developed over the last decade at the Raman Research Institute and optimized to detect the global 21~cm signal in the $50-200$~MHz range, i.e. in the Cosmic Dawn and Epoch of Reionization.
The SARAS antennas have been designed to provide nearly frequency independent beams and avoid coupling of sky spatial structures into spectral structures in order to preserve the intrinsically smooth foreground spectrum (\cite{sathyanarayana17}).
Initially, SARAS featured a fat-dipole antenna (\cite{patra13}) that was later replaced by a shaped monopole antenna (SARAS~2, \cite{singh18a}). SARAS~2 was deployed in the radio-quiet Timbaktu Collective in Southern India, observing in the $110-200$~MHz. SARAS~2 results disfavoured models with inefficient heating of the intergalactic medium and rapid reionization (\cite{singh17}, \cite{singh18b}). 
A new generation experiment, SARAS~3, exploits a refined design to further reject spurious foreground structures and control over systematics in order to target the 21~cm signal in the $50-100$~MHz band.

\section{Space-based instruments}
\label{sec:space_base_instruments}

Whereas tremendous progress is being made from the ground to detect the globally-averaged and spatially-fluctuating 21-cm signal during the EoR and CD, as discussed earlier, the stability of the system, RFI, the ionosphere, and even multi-path propagation effects, make ground-based observations hard and in some cases, such as a detection of the Dark Ages, even impossible. These motivations have been driving concepts and plans for space-based instrumentation. 
%Are view of some of these can be founds in [Koopmans et al. (2019)]

\subsection{The Dark Ages Polarimetry Pathfinder -- DAPPER}

The Dark Ages Polarimetry Pathfinder (DAPPER, \cite{burns19}) is a space satellite that is intended to observe the global signal from a $\sim 50000$~km lunar orbit, one of the quietest radio frequency environments, with an expected 26~month lifetime. Its goal is to observe the global signal absorption trough expected at $17 < \nu < 38$ ($83 > z > 36$), e.g. in the Dark Ages, well before the formation of the first luminous sources. In this epoch, the global signal is determined by linear perturbation theory, uncontaminated by complex astrophysical processes. DAPPER is expected to characterize the predicted signal, including any deviation that may be due by the additional cooling reported by \cite{bowman18}. Its strategy includes the use of a polarimeter to measure polarization induced by the anisotropic foregrounds, a large antenna beam to aid the foreground separation from the isotropic, unpolarized global signal (\cite{nhan17}) and a pattern recognition data analysis that is trained on realisti smulations of observed foregrounds, instrument systematics and the expected global signal (\cite{tauscher18}).
DAPPER is one of nine small satellite missions selected by NASA to be further studied for a possible launch in the next decade.

\subsection{Discovering the Sky at the Longest Wavelengths -- DSL}

The Discovering the Sky at the Longest Wavelengths (DSL, \cite{chen19}) is a mission concept that explores the possibility to deploy a constellation of micro-satellites circling the Moon on nearly-identical orbits, performing interferometric observations of the sky below 30~MHz. 
Although its sensitivity is insufficient to detect 21~cm fluctuations from the Dark Ages, its goals will be to accurately image 21~cm foregrounds and to target the 21~cm global using a calibrated single antenna. 
The current DSL concept includes a larger "mother" satellite that leads or trails $5-8$ smaller daughter satellites that carry out the radio observations and pass the data to the mother satellite through a microwave link. The mother performs the cross correlation and handles communications with the Earth. 
The DSL project is now undergoing a prototype study.

\subsection{Farside Array for Radio Science Investigations of the Dark ages and Exoplanets -- FARSIDE}

The Farside Array for Radio Science Investigations of the Dark ages and Exoplanets (FARSIDE, \cite{burns19b}) is a mission concept to place an interferometric array on the far side of the Moon, which offers complete isolation from terrestrial radio frequency interference and solar wind, allowing observations at sub-MHz frequencies. The array would consist of 128 dual polarization antennas deployed across a 10~km area by a rover, observing in the $0.1-100$~MHz (basically $z > 13$) range. FARSIDE would also include precision calibration of an individual antenna element via an orbiting beacon in order to attempt the detection of the global 21~cm from the Dark Ages ($50 < z < 100$).
A NASA-funded design study, focused on the instrument, a deployment rover, the lander and base station, delivered an architecture broadly consistent with the requirements for a Probe mission (about 1.3 billion USD).

\subsection{Netherlands-China Low frequency Explorer -- NCLE}

The Netherlands-China Low frequency Explorer (NCLE) is a radio instrument payload on board on the Chinese Queqiao relay satellite that orbits behind the Moon. 
%(Fig.~\ref{fig:fig_NCLE}; https://www.ru.nl/astrophysics/radboud-radio-lab/projects/netherlands-china-low-frequency-explorer-ncle/). 
NCLE is designed, built and tested by a Dutch consortium comprised of the Radboud University, ASTRON and ISIS, in close collaboration with the National Astronomical Observatories of the Chinese Academy of Sciences. It is composed of three, 5~m-long, carbon-fibre monopole antenna units that can be switched into dipole mode to observe in the $0.08-80$~MHz range. Its main target is therefore the global 21~cm signal from the Darks Ages and the Cosmic Dawn although it will also provide accurate, degree-scale foreground maps below 10~MHz and an extensive characterization of the radio frequency interference environment in the lunar far side.
The Queqiao satellite was launched on May 21st, 2018, and is currently behind the Moon, in the Earth-Moon second Lagrange point. NCLE is currently being commissioned, with first observations starting before the end of 2019.
%\begin{figure}[]
%\begin{center}
%\includegraphics[width=0.6\textwidth]{Koopmans_Bernardi/NCLE_AJDI_edited.jpg}
%\end{center}
%\caption{A rendering of the Netherlands-China Low-frequency Explorer (NCLE), currently in lunar L2. [ASTRON-JIVE Daily Image]}
%\label{fig:fig_NCLE}
%\end{figure}

\section{The far future of 21-cm cosmology}

The 21-cm signal instruments and experiments described in this book and chapter often have continuously operated already for close to decade (e.g. EDGES, LOFAR, MWA) and have made tremendous progress, possible being on the verge of a detection or enabling to exclude a wide range of 21-cm signal models, either standard (\cite{monsalve17b}, \cite{singh17}, \cite{monsalve18}, \cite{singh18b}) or "exotic" (\cite{barkana18}, \cite{fialkov18}, \cite{fialkov19}). Some have already been decommissioned and/or are being upgraded since they are reaching the end of their physical lifetime or are merely reaching the maximum of their capabilities (being thermal-noise or systematic-error limited). Entirely new ground-based instruments are also coming online or are being designed at the moment (e.g. NenuFAR, HERA, SKA) to push boundaries various parameter spaces and which will likely dominate 21-cm signal science in the coming decade or two. 

{\sl However, what lies beyond these instruments? What is still "left to do" when those future instruments have maximised their science return?} 

As earlier touched upon, besides pushing the boundaries of parameter space in redshift, spatial scale and signal-to-noise (e.g. imaging versus power-spectra measurements), most ground-based instrument are running or will run against limits due to human-made RFI and ionospheric errors at very low frequencies. 

The penultimate 21-cm signal instrument should, therefore, be in space, away from RFI and ionospheric errors, enabling not only extremely precise and accurate measurements of the 21-cm signal covering the redshifts of the EoR and CD, but ultimately make a detection of 21-cm signal from the Dark Ages, and also enable direct imaging of the Cosmic Dawn. The challenges that are facing ground-based instruments are impossible to overcome to reach those objectives (\cite{koopmans19}). For example, at frequencies corresponding to Dark Ages redshifts, any radio signal from the sky is almost 100\% distorted by the ionosphere on time-scales of seconds with variable distortions over arcminutes scales. It is therefore impossible to correct for these errors and space instruments will be needed in order to observe the Dark Ages and image the Cosmic Dawn.

Space instruments are, however, complex and expensive and developing a light-weight, durable and space-proof space technology is therefore critical. Moreover, space environment is not ubiquituosly RFI free, like Earth orbits. An optimal location is either in deep space where the earth becomes a faint radio source that can be dealt with using traditional excision techniques - but where the sun remains a source of noise -, the back-side of the moon (e.g. FARSIDE, \cite{burns19b}) or in lunar orbit (e.g. DSL, DAPPER, \cite{chen19}, \cite{burns19}), where it will be shielded from both the earth and the sun for a fraction of time.
%Even NCLE being about 60,000\,km behind the moon in L2, is not optimally located. With increasing lunar activity (e.g. landers, orbiters), however, even RFI from the lunar surface or orbit might become an issue of concern at some point in time if not carefully thought through or shielded. 
It should also be kept in mind that the lunar surface is partly charged due to solar radiation and cosmic rays and hence even on the lunar surface effects of a lunar "ionosphere" are not completely absent. On the other hand, reflections of radio waves from the lunar surface back to any orbiter could lead to multi-path propagation and also need mitigation since the dynamic range of the signal can be as high as $10^8$ for the Dark Ages (a $10^5$~K foreground sky versus a mK signal).
  
Besides these ``environmental" effects, any space-based interferometer should have a collecting area far exceeding the area of upcoming instruments like HERA and the SKA. 
%This requirement is not there for global-signal experiments where signal-to-noise is independent of the size of the instrument (of course placing say a hundred similar instruments in space will reach the desired global signal ten times faster). 
As the number of visibilities per spatial or $uv$-resolution element and their thermal noise determines the power spectrum sensitivity, future instruments will feature an increased number of receiving elements that are cross-correlated (therefore measuring more modes) together with a larger collecting area (measuring each model with more sensitivity). \cite{koopmans19} suggests that the collecting area necessary to observe the Dark Ages will be as large as $10-100$~km$^2$ in order to overcome the sky-dominated noise from foregrounds that are $\sim 10^5$~K bright. Despite these challenges, the information contained in the 21-cm signal during the Dark Ages and the early Cosmic Dawn will shed light on fundamental physics as well on the astrophysics of the infant Universe, making these developments worth the effort.

\bibliographystyle{plain}
\bibliography{Koopmans_Bernardi/References}

%\part{Theory}
%\appendix

\end{document}